\documentclass[aps%
,pre%
,showpacs%
,twocolumn%
,superscriptaddress%
,floats%
,amssymb%
]{revtex4}

\usepackage{graphicx}
\usepackage{amsmath}
\usepackage{amssymb}
\bibstyle{apsrev.bib}

\newcommand{\comment}[1]{}
\begin{document}

\title{Exact Enumeration of Ground States in the Sherrington-Kirkpatrick
Spin Glass}
\date{\today{}}
\author{Stefan Boettcher}
\email{sboettc@emory.edu}
\affiliation{Physics Department, Emory University, Atlanta, Georgia 30322, USA}
\author{Tomasz M. Kott}
\email{tkott@bucknell.edu}
\affiliation{Physics Department, Bucknell University, Lewisburg, Pennsylvania
17837, USA}

\begin{abstract}
Using the discrete $\pm J$ bond distribution for the
Sherrington-Kirkpatrick spin glass, all ground states for the entire
ensemble of the bond disorder are enumerated. Although the
combinatorial complexity of the enumeration severely restricts
attainable system sizes, here $N\leq9$, some remarkably intricate
patterns found in previous studies already emerge. The analysis of the
exact ground state frequencies suggests a direct construction of their
probability density function. Against expectations, the result
suggests that its highly skewed appearance for finite $N$ evolves
logarithmically slow towards a Gaussian distribution.
\end{abstract}
\maketitle

The Sherrington-Kirkpatrick (SK) model~\cite{Sherrington75} of glassy
behavior in magnetic materials has provided a conceptual framework for
the effect of disorder and frustration that are observed in systems
ranging from materials~\cite{F+H} to combinatorial optimization and
learning~\cite{MPV}.  It's conceptual simplicity is expressed through
the Hamiltonian
\begin{eqnarray}
H=\frac{1}{\sqrt{N}}\sum_{i<j}^{N}J_{i,j}\sigma_{i}\sigma_{j},
\label{Heq}
\end{eqnarray}
in which all pairs of binary Ising spin-variables $\sigma_i=\pm1$ are
mutually connected through a bond matrix $J_{i,j}$, which is symmetric
and whose entries are random variables drawn from a distribution
$P(J)$ of zero mean and unit variance. We note that this Hamiltonian
possesses a local ``gauge''-invariance under the transformation of
\begin{eqnarray}
\sigma_i\to -\sigma_i\quad{\rm and}\quad J_{i,j}\to -J_{i,j},
\label{gaugeeq}
\end{eqnarray}
at any site $i$ and the bonds to all its adjacent sites
$j$~\cite{Toulouse77}.

The SK model has reached significant prominence because, despite of
its apparent simplicity, its solution proved surprisingly difficult,
revealing an amazing degree of complexity in its structure~\cite{MPV}.
While it is solvable in principle, many of its features have not been
derived yet. One such feature concerns the probability density
function (PDF) of its ground state energies. Being an extreme element
of the energy spectrum, the distribution of $e_0$ is not necessarily
normal but instead may follow a highly skewed ``extreme-value
statistics'' as can be derived for the Random Energy
Model~\cite{Bouchoud97}. If the energies within that spectrum are
uncorrelated, it can be shown that the PDF for $e_{0}$ is among one of
only a few universal functions. This extreme-value statistics of the
ground states has been pointed out in Ref.~\cite{Bouchoud97} and has
received considerable attention
recently~\cite{Andreanov04,Palassini03,EOSK,Katzgraber05}.  For
instance, if the sum for $H$ in Eq.~(\ref{Heq}) were over a large
number of independent terms, $H$ would be Gaussian distributed.  In
such a spectrum, the probability of finding $H\to-\infty$ decays
faster than any power, and ground states $e_{0}$ should be distributed
according to a Gumbel PDF,~\cite{Bouchoud97}
\begin{eqnarray}
g_{m}(x)\propto\exp\left\{m\left(x-e^x\right)\right\}
\label{Gumbeleq}
\end{eqnarray}
with $m=1$, here generalized to the case where $m$ refers to the
$m$-th lowest extreme value~\cite{Palassini03}.

\begin{figure}
\vskip 2.4in
\includegraphics{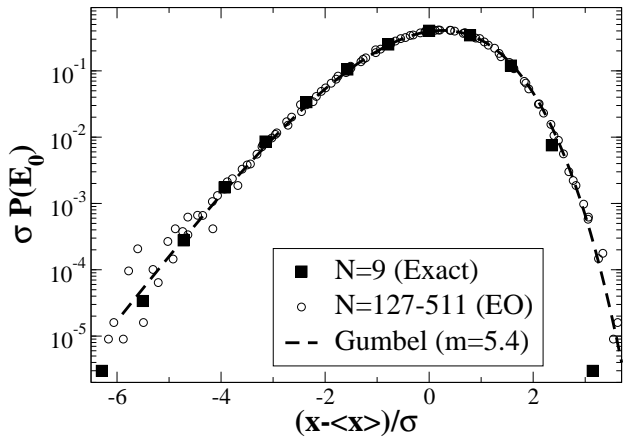}
\caption{Plot of the rescaled probability density function of ground
state energies using $\pm J$ bonds. Shown are the numerical data for
the SK model obtained with the extremal optimization (EO) heuristic
and their fit by Eq.~(\protect\ref{Gumbeleq}) with $m=5.4$, properly
rescaled, from Ref.~\protect\cite{EOSK}, and the exact results for
$N=9$ based on Tab.~\protect\ref{energylevels}.  Remarkably, there is
very little variation between the exact data at $N=9$ and the
numerical data obtained for $N=127-511$.
}
\label{pdfplot}
\end{figure}

In a spin glass the individual terms in Eq.~(\ref{Heq}) are
not independent variables and deviations from any universal behavior
may be expected. In particular, these deviations should become
strongest when all spin variables are mutually interconnected such as
here in the SK model, but may be less so for sparse graphs, such as
low-dimensional lattices. (Although it should be noted that sparsely
connected systems seem to have a Gaussian PDF, see Ref.~\cite{EOSK},
provably so finite-dimensional lattices~\cite{Wehr90}.) Indeed, in
mean-field models
Refs.~\cite{Andreanov04,Palassini03,Bouchaud03,Katzgraber05} find
numerically a highly skewed PDF for $e_{0}$ which does not fit to the
Gumbel distribution in Eq.~(\ref{Gumbeleq}) for an $m=1$-lowest value. In
Fig.~\ref{pdfplot}, we plot the rescaled PDF of ground state energies
in the SK obtained in Ref.~\cite{EOSK} for $\pm J$ bonds on systems of
size $N=127-511$ with the extremal optimization (EO)
heuristic~\cite{Boettcher01a}. The result resembles that of
Ref.~\cite{Palassini03} to a surprising degree.  In fact, a naive fit
of Eq.~(\ref{Gumbeleq}) for variable $m$ to the SK-data, as suggested
by Ref.~\cite{Palassini03}, yields virtually identical results, with
$m\approx5$.

In this Communication, we derive the PDF in Eq.~(\ref{Gumbeleq}) {\it
analytically,} motivated from the study of exact enumerations of ground
states in the SK for small $N$. We find an $N$-dependent parameter
\begin{eqnarray}
m\sim\ln(N)
\label{meq}
\end{eqnarray}
to leading order.  As $m$ grows with $N$, the
$g_m(x)$ ultimately develops a (symmetric)
Gaussian form. Yet, $m$ grows sufficiently weakly to justify the the highly skewed appearance of the PDF observed
numerically over a wide range of sizes $N$, see
Fig.~\ref{pdfplot}.

Unlike for the Gaussian bond distribution, discrete bonds allow a
complete enumeration of the ground-state PDF for each possible bond
structure at small system sizes $N$.  The motivation for such a study
is provided by the hope to discern certain patterns in the solutions
that may be extrapolated to the large-$N$ limit. Looking at the PDF as
a whole, we find at small $N$ a highly skewed function. In
Fig.~\ref{pdfplot} we show that the exact result for $N=9$ compares
quite well with the numerical sampling at larger $N$, and with the
Gumbel fit. A detailed analysis of the results at $N\leq9$ suggests a
{\it direct} construction of the PDF assuming only local gauge
symmetry.

For the $\pm J$ bond distribution, complete enumeration entails a
generation of all possible symmetric $N\times N$ bond matrices
$J_{i,j}$, filled with all combinations of $J_{i,j}=+J$ or
$-J$. (Diagonal elements, corresponding to self-coupling terms, are
zero.) For each instance, we then determine the ground state. Clearly,
the combinatorial effort involved in merely generating all instances
becomes prohibitive already for small $N$, restricting us here to
$N\leq9$. For $N=9$ the program runs for about 5h on a 2GHz
computer. Larger $N$ within reasonable CPU-time probably could have
been reached with a more elaborate algorithm that recursively
constructs from $N-1$ to $N$ only distinct instances by adding a new
vertex with all possible bond combinations, and determines their
ground state and their weight. We merely exploit the fact that we can
always use Eq.~(\ref{gaugeeq}) to gauge-transform to have one spin in
any instance to, say, only possess $J=+1$ bonds, leaving us with
$(N-1)(N-2)/2$ variable bonds or $2^{(N-1)(N-2)/2}$ instances at each
$N$.

\begin{table}
\caption{Count $n_{B}(N)$ of the number of instances of size $N$
having $B$ violated bonds in the ground state. Rescaled as a PDF,
$n_{B}(9)$ is plotted in Fig.~\protect\ref{pdfplot}.}
\begin{tabular}{c|r|r|r|r|r|r|r}
\hline 
$N$& 3& 4& 5& 6& 7& 8& 9\tabularnewline
\hline
$n_{0}$& 1& 1& 1& 1& 1& 1& 1\tabularnewline
 $n_{1}$& 1& 6& 10& 15& 21& 28& 36\tabularnewline
 $n_{2}$& & 1& 30& 105& 210& 378& 630\tabularnewline
 $n_{3}$& & & 22& 395& 1\,260& 3\,276& 7\,140\tabularnewline
 $n_{4}$& & & 1& 480& 4\,830& 20\,195& 58\,590\tabularnewline
 $n_{5}$& & & & 27& 11\,382& 92\,232& 367\,668\tabularnewline
 $n_{6}$& & & & 1& 11\,322& 308\,756& 1\,814\,358\tabularnewline
 $n_{7}$& & & & & 3\,720& 702\,376& 7\,118\,100\tabularnewline
 $n_{8}$& & & & & 21& 787\,787& 22\,106\,430\tabularnewline
 $n_{9}$& & & & & 1& 180\,036& 52\,700\,060\tabularnewline
 $n_{10}$& & & & & & 2\,058& 84\,901\,278\tabularnewline
 $n_{11}$& & & & & & 28& 72\,434\,628\tabularnewline
 $n_{12}$& & & & & & 1& 25\,335\,810\tabularnewline
 $n_{13}$& & & & & & & 1\,590\,060\tabularnewline
 $n_{14}$& & & & & & & 630\tabularnewline
 $n_{15}$& & & & & & & 36\tabularnewline
 $n_{16}$& & & & & & & 1 \tabularnewline
\hline
\end{tabular}
\label{energylevels}
\end{table}

In Tab.~\ref{energylevels} we list the count $n_B(N)$ for instances
with $B$ violated bonds, starting from the ferromagnetic (FM) instance
with $B=0$ to the totally anti-ferromagnetic (AFM) instance with
\begin{eqnarray}
B=B_{\rm max}=
\begin{cases}
(N-1)^2/4,& N $~odd$,\cr N(N-2)/4,& N $~even$.
\end{cases}
\label{Bmaxeq}
\end{eqnarray}
For instance, after accounting for the gauge symmetry, $n_{0}(N)=1$
corresponds to the one instance with perfect ferromagnetic order,
while on the bottom of each column we find the one perfect
anti-ferromagnet with the largest number of violated bonds $B_{\rm
max}$, which arises because the energy of the AFM is minimized when up
and down spins divide as evenly as possible (still leaving about half
of all ${N\choose2}$ bonds violated).  These numbers for $N=9$,
properly rescaled, are plotted in Fig.~\ref{pdfplot}.

In an attempt to discern a pattern in the growth of these numbers, we
found a few interesting facts. Starting for a given $N$ at the FM
instance ($B=0$), we find
\begin{eqnarray}
n_B(N)={{N\choose2}\choose B},\qquad
\left(B<\left\lfloor\frac{N}{2}\right\rfloor\right),
\label{ferroeq}
\end{eqnarray}
where $\lfloor x\rfloor$ refers to the next-smallest integer to $x$.
In this regime, the $n_B(N)$ instances with $B$ bond violations are
due to all possible embeddings of $B$ independent $-J$ bonds in the
fully connected graph of size $N$. FM order is preserved and each new
$-J$ bond increments the number $B$ of violations. At $B=\lfloor
N/2\rfloor$, a few embeddings, where all $B$ of such $-J$ bonds are
connected to the same spin, can be gauge-transformed into instances of
lower ($N$ even) or equal ($N$ odd) number of violations, {\it
reducing} the number of independent instances below the prediction
Eq.~(\ref{ferroeq}). In general, for further increasing $B$, more and
more $-J$ bonds can be gauge-transformed away at spins where $-J$
bonds equal or outnumber $+J$ bonds. A similar behavior is observed
starting with the AFM instance: initially each $+J$ bond added serves
to {\it decrement} the number of violated bonds without changing the
AFM order. This leads to $n_{B_{\rm max}-i}(N)$ just as given in
Eq.~(\ref{ferroeq}), but only for $i\leq\lfloor
(N-5)/2\rfloor$. Instances with $i=\lfloor (N-3)/2\rfloor$ fewer bond
violations than the AFM are suddenly far {\it more} numerous,
consisting not only of those with $i$ embedded $+J$ bonds, but also
certain instances with $i+2$ such bonds, in particular $(i+2)$-gons
and -trees. While in the AFM state for a given $N$ half of the spins
are up, the other half down, adding $i+2=\lfloor (N+1)/2\rfloor$ {\it
connected} $+J$ bonds must link oppositely oriented spins, which does
not help to reduce violated bonds.

Finding a closed form expression for the number of instances with a
certain number of bond violations beyond these limits near the FM and
the AFM instances appears difficult. But we can venture to make some
approximations to capture the noticeable asymmetry in the PDF visible
in Fig.~\ref{pdfplot}. As mentioned above, Eq.~(\ref{ferroeq}) breaks
down for larger numbers of $-J$ bonds because instances with any spins
having equal or more than half their bonds being $-J$ can be gauge
transformed and may have to be discounted. A conceivable approximation
than is to {\it only} count instances in which no spin has more than
$\lfloor (N-1)/2\rfloor$ of $-J$ bonds, rejecting instances that could
be gauge-transformed into other instances satisfying this condition.
This approximation reproduces Eq.~(\ref{ferroeq}) exactly, {\it and}
it predicts exactly that there are no instances with more than $B_{\rm
max}$ violated bonds as given in Eq.~(\ref{Bmaxeq}).

Based on these arguments, we now construct an analytic expression for
$n_B(N)$ for $N\to\infty$. As above, we consider embedding a random
graph~\cite{Bollobas} of $-J$ bonds in a fully-connected graph
otherwise filled with $+J$ bonds. With the above condition, we assume
that in each instance the number of $-J$ bonds corresponds to the
number of violations $B$, which is not generally true. An added bond
gives each spin a probability of $2/N$ to be linked to it. Then, the
probability of a single spin having a degree $d$ of $-J$ bonds is
$p_d={B\choose d}(2/N)^d(1-2/N)^{B-d}$. Hence, the probability for a
single spin to have a degree $d<d_{\rm max}=\lfloor (N-1)/2\rfloor\sim
N/2$ is
\begin{eqnarray}
p_{d<d_{\rm max}}\sim\sum_{d=0}^{N/2}{B\choose
d}\left(\frac{2}{N}\right)^d\left(1-\frac{2}{N}\right)^{B-d}.
\label{probeq}
\end{eqnarray}
Naively assuming that each spin can be treated independently, an
instance as a whole has a probability of $p_{d<d_{\rm max}}^N$ to be
counted. For the total number of instances with $B$ violations, we
obtain
\begin{eqnarray}
n_B(N)\approx {{N\choose2}\choose B}~p_{d<d_{\rm max}}^N.
\label{counteq}
\end{eqnarray}

Asymptotic analysis for $N\to\infty$ with $B=N^2/4(1+v)$, $v\ll 1$,
shows that ${{N\choose2}\choose B}\sim\exp\{-N^2v^2/4\}$ at its peak.
An similar evaluation of $p_{d<d_{\rm max}}$ in Eq.~(\ref{probeq}) for
$1\ll N\ll B\leq N^2/2$ yields an internal saddle point~\cite{BO} at
$d\sim2B/N$ for $B\ll N^2/4$ where $p_{d<d_{\rm max}}\sim1$ with
exponentially small corrections, whereas for $N^2/4\leq B\leq N^2/2$
the sum in Eq.~(\ref{probeq}) can be evaluated at its upper limit,
$d=N/2$, to yield
\begin{eqnarray}
p_{d<d_{\rm max}}&\sim&\exp\{-N/2[4B/N^2-1-\ln(4B/N^2)]\}\\
&\sim&\exp\{-N/2[v-\ln(1+v)]\}.
\label{asympeq}
\end{eqnarray}
Note that in the last exponential the linear term in $v$ {\it exactly
cancels} for $v\ll1$. In that case, $p_{d<d_{\rm
max}}\sim[1+\exp(vN/2)]^{-v/2}$ provides an interpolation for both of
its asymptotic regimes. Then, with $v=2w/N$, we obtain $n_B(N)\propto
e^{-w^2}/(1+e^{w})^{w}$ in Eq.~(\ref{counteq}). This result is indeed
independent of $N$ but a very poor (essentially Gaussian) fit of the data.

On the other hand, we may suppose that the cancellation is accidental,
due to the neglect of nontrivial correlations in the above discussion,
and assume that the more generic result in Eq.~(\ref{asympeq}) is
simple exponential, $\exp(-\alpha vN/2)$ (with some $\alpha>0$),
leading to an interpolation
\begin{eqnarray}
p_{d<d_{\rm max}}\sim1+\exp(\alpha vN/2).
\label{interpolationeq}
\end{eqnarray}
With $v=2w/N$, we obtain
\begin{eqnarray}
n_B(N)\propto \frac{e^{-w^2}}{(1+e^{\alpha w})^N}.
\label{newBeq}
\end{eqnarray}
This expression for $N\to\infty$ has a {\it moving} saddle
point~\cite{BO} at
\begin{eqnarray}
w_0\sim-\frac{1}{\alpha}\left[\ln\left(\alpha\frac{N}{2}\right)-\ln\ln\left(\alpha\frac{N}{2}\right)+\frac{\ln\ln\left(\alpha\frac{N}{2}\right)}{\ln\left(\alpha\frac{N}{2}\right)}\right],
\label{saddleeq}
\end{eqnarray}
listing all orders in the expansion needed to facilitate a stationary
saddle point.  Transforming onto the saddle point by substituting
$w=w_0+x/\alpha$ into Eq.~(\ref{newBeq}) and expanding in $x\ll w_0$
(i.~e. $x\ll\ln N$, so the $x$-domain spans the entire real line for
$N\to\infty$), we finally obtain Eq.~(\ref{Gumbeleq}) with an
$N$-dependent parameter $m=-2w_0/\alpha$ which reduces to
Eq.~(\ref{meq}).

Finally, we can use the data in Tab.~\ref{energylevels} also to
consider the average number of violated bonds in the ground state. For
each $N$, this average is a rational number, the denominator being the
total number of instances in the ensemble. Thus, the numerator
provides a sequence of integers that is simply related to the average
ground state energy.  Tab.~\ref{averages} shows the moments of the
energy distribution. To retain integer numbers, we define
\begin{eqnarray}
{\mathcal{E}}_{B}(N)= 2B-{N\choose2},
\end{eqnarray}
which is related to the true energy of an instance by
$E={\mathcal{E}}_B(N)/\sqrt{N}$, according to Eq.~(\ref{Heq}). Then
one would expect that
$\lim_{N\to\infty}\langle{\mathcal{E}}(N)\rangle/N^{3/2}\approx-0.7633$,
the Parisi solution for the ground state energy density of the SK
model~\cite{MPV}. Here, $\langle{\mathcal{E}}(N)\rangle$ is the ratio
of the value in the third and the second column in Tab.~\ref{averages}
for each $N$. While the denominator is simply the number of instances,
$I=2^{N-1\choose2}$, the value in the third column is highly
nontrivial with huge prime-factors. Even a partial identification of
that sequence of numbers seems intractable.

\begin{table}
\caption{List of the number of instances $I$ and the first two moments
(unnormalized) of the distributions listed in
Tab.~\protect\ref{energylevels}}
\begin{tabular}{c|r|r|r}
\hline 
$N$&$I$&$\sum_{B}{\mathcal{E}}_{B}n_{B}(N)$&$\sum_{B}{\mathcal{E}}_{B}^{2}n_{B}(N)$\tabularnewline
\hline 
3 & 2 & -4 & 10\tabularnewline 
4 & 8 & -32 & 136\tabularnewline
5 & 64 & -360 & 2\,176\tabularnewline 
6 & 1\,024 & -8\,418 & 71\,664\tabularnewline 
7 & 32\,768 & -338\,928 &3\,645\,600\tabularnewline 
8 & 2\,097\,152 & -28\,189\,776 &388\,069\,088\tabularnewline 
9 & 268\,435\,456 & -4\,294\,748\,800 &70\,448\,532\,736
\tabularnewline 
\hline
\end{tabular}
\label{averages}
\end{table}

SB thanks H. Katzgraber and A. Hartmann for helpful discussions.
This work has been partially supported by grant \#0312510 from the
Division of Materials Research at the National Science Foundation. TK
thanks the REU program at the National Science Foundation and the SURE
program at Emory University for its support. 

\comment{
These commands specify that your bibliography should be created from entries
in the file bibdata.bib. There are several bibliography styles that can be
used with the eethesis document class; the style theunsrt.bst is the style
used in the examples in this manual. The entries in this style are patterned
after those in IEEE Transactions on Automatic Control. It lists the sources
in the order they were cited. There is also a generic ieeetr.bst, which
formats sources similar to many IEEE publications. If neither of these
styles is suitable for your department, you might consider acm.bst or
siam.bst which format your bibliography in the style of ACM and SIAM
publications. Check the /usr/local/teTeX/local.texmf/bibtex/bst directory on
the machine that you use to see if there are any other .bst files you can
use.}
\bibliographystyle{apsrev}
\bibliography{exactSK3}

\end{document}